\documentclass{IEEEoj}
\usepackage{cite}
\usepackage{amsmath,amssymb,amsfonts}
\usepackage{algorithmic}
\usepackage{graphicx,color}
\usepackage{textcomp}
\usepackage{cite}
\usepackage{amsmath,amssymb,amsfonts}
\usepackage{hyperref}
\usepackage{algorithmic}
\usepackage{graphicx}
\usepackage{textcomp}
\usepackage{xcolor}
\usepackage{array}
\usepackage[acronym]{glossaries}
\usepackage{colortbl}
\usepackage{todonotes}
\setuptodonotes{inline}
\usepackage{booktabs}
\usepackage{multirow}
\usepackage{layouts}
\usepackage{tikz, adjustbox}
\usepackage[most]{tcolorbox}
\usepackage{wrapfig}
\usepackage{subcaption}
\usepackage{amsmath,amsfonts}
\usepackage{algorithmic,algorithm}
\usepackage{subcaption}
\usepackage{textcomp}
\usepackage{stfloats}
\usepackage{url}
\usepackage{verbatim}
\usepackage{graphicx}
\usepackage{pifont}
\usepackage{bbding}
\usepackage{colortbl}
\usetikzlibrary{patterns}
\usepackage{pgfplots}

\usepackage{todonotes}
\usepackage{tikz}
\usepackage{xfp}
\usepackage{blindtext}

\usepackage{tikz}
\usepackage{amsmath}
\usepackage{array}
\usepackage{tabularx}

\newacronym{PQC}{PQC}{Post-Quantum Cryptography}
\newacronym{PQ}{PQ}{Post-Quantum}
\newacronym{DoS}{DoS}{Denial-of-Service}
\newacronym{DLP}{DLP}{discrete logarithm problem}
\newacronym{KEM}{KEM}{Key Encapsulation Mechanism}
\newacronym{SIS}{SIS}{Shortest Integer Solution}
\newacronym{LWE}{LWE}{Learning With Errors}
\newacronym{CT}{CT}{Certificate Transparency}
\newacronym{PKI}{PKI}{Public Key Infrastructure}
\newacronym{AP}{AP}{Access Point}
\newacronym{QC}{QC}{quantum computing}
\newacronym{EAP}{EAP}{Extensible Authentication Protocol}
\newacronym{NIST}{NIST}{National Institute of Standards and Technology}
\newacronym{MITM}{MITM}{Man-in-the-Middle}
\newacronym{NTRU}{NTRU}{Number Theorists 'R' Us}
\newacronym{PTK}{PTK}{Pairwise Transient Key}
\newacronym{PMK}{PMK}{Pairwise Master Key}
\newacronym{EAP-TTLS}{EAP-TTLS}{Extensible Authentication Protocol-Tunneled Transport Layer Security}
\newacronym{EAP-TLS}{EAP-TLS}{Extensible Authentication Protocol-Transport Layer Security}

\usepackage[dvipsnames]{xcolor}

\definecolor{hses-primary}{HTML}{193058}
\definecolor{hses-secondary}{HTML}{b6163d}
\definecolor{hses-tertiary}{HTML}{00aadc}
\definecolor{hses-background}{HTML}{f0f0f0}

\definecolor{hses1-primary}{HTML}{002a58}
\definecolor{hses1-secondary}{HTML}{D50539}
\definecolor{hses1-tertiary}{HTML}{01aadb}
\definecolor{hses1-background}{HTML}{f0f0f0}

\definecolor{brew-primary}{HTML}{f1a340}
\definecolor{brew-secondary}{HTML}{998ec3}
\definecolor{brew-tertiary}{HTML}{f7f7f7}

\definecolor{drawio-green-1}{HTML}{82B366}
\definecolor{drawio-green-2}{HTML}{D5E8D4}
\definecolor{drawio-red-1}{HTML}{B85450}
\definecolor{drawio-red-2}{HTML}{F8CECC}
\definecolor{drawio-blue-1}{HTML}{6C8EBF}
\definecolor{drawio-blue-2}{HTML}{DAE8FC}
\definecolor{drawio-yellow-1}{HTML}{D6B656}
\definecolor{drawio-yellow-2}{HTML}{FFF2CC}
\definecolor{drawio-orange-1}{HTML}{FFE6CC}
\definecolor{drawio-orange-2}{HTML}{D79B00}

\definecolor{OrchidGreen}{HTML}{A5CE5E}
\definecolor{PastelGreen}{HTML}{96DD99}
\definecolor{PastelRed}{HTML}{FF6962}

\definecolor{common}{HTML}{000000}
\definecolor{uncommon}{HTML}{6D9EEA}
\definecolor{rare}{HTML}{32CD32}
\definecolor{legendary}{HTML}{EFBF04}
\definecolor{mythic}{HTML}{FF00FF}

\definecolor{24ghz}{HTML}{808080}
\definecolor{5ghz}{HTML}{FFFFFF}

\definecolor{ap}{HTML}{627C87} 
\definecolor{client}{HTML}{939393}
\definecolor{radius}{HTML}{6B8653}

\definecolor{ap5}{RGB}{164,206,225}
\definecolor{client5}{RGB}{245,245,245}
\definecolor{radius5}{RGB}{179,224,139}

\usepackage{amssymb}
\usepackage{fontawesome5}
\usepackage{wasysym}
\usepackage{tikz}

\newcommand{\mailto}[1]{
\href{mailto:#1}{\tikz[baseline=(char.base)]{\node[shape=circle,draw=black,fill=black,inner sep=2pt] (char) {\textcolor{white}{\fontsize{6}{8}\faIcon[regular]{envelope}}}}}
}

\newcommand{\github}[1]{\footnote{\faIcon{github}~\url{#1}}}

\newcommand{\orcidicon}{\includegraphics[width=0.32cm]{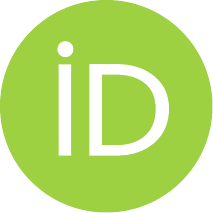}}
\newcommand{\orchid}[1]{\href{https://orcid.org/#1}{\orcidicon}}

\usepackage{calc} 

\newlength{\labelwidthlen}

\newlength{\barwidthpt}
\newlength{\linewidthpt}

\newcommand{\stackedbar}[4]{%

\ifnum#4=1
    \pgfplotsset{bandstyle/.style={}}
    \def\colA{client}%
    \def\colB{radius}%
    \def\colC{ap}%
\else
    \pgfplotsset{bandstyle/.style={pattern=north east lines, pattern color=}}
    \def\colA{client5}%
    \def\colB{radius5}%
    \def\colC{ap5}%
\fi

\begin{tikzpicture}
    \begin{axis}[
        hide axis,
        xbar stacked,
        width=\linewidth,
        height=0.28cm,
        xmin=0, xmax=#1 + #2 + #3,
        axis line style={draw=none},
        tick style={draw=none},
        xtick=\empty, ytick=\empty,
        enlarge y limits=false,
        bar width=0.28cm,
        scale only axis,
        axis on top=false,
    ]

    \newcommand{\formatval}[1]{%
        \pgfmathparse{round(##1*100)/100}%
        \pgfmathprintnumberto[fixed, precision=2]{\pgfmathresult}{\formatted}%
        \formatted
    }

    \newcommand{\maybeLabel}[3]{%
      \sbox{\tempbox}{\scriptsize\formatval{##1}}%
      \setlength{\labelwidthlen}{\wd\tempbox}%

      \pgfmathsetmacro{\ratio}{##1/(##1+#2+#3)}%
      \setlength{\linewidthpt}{\linewidth}%
      \setlength{\barwidthpt}{\ratio\linewidthpt}%

      \ifdim \labelwidthlen < 0.7\barwidthpt 
      \node[anchor=center, ##3] at (axis cs:##2,0) {\scriptsize\formatval{##1}};
      \fi
    }

\ifnum#4=1
    \maybeLabel{#1}{#1/2}{white}
    \maybeLabel{#2}{#1 + #2/2}{white}
    \maybeLabel{#3}{#1 + #2 + #3/2}{white}
\else
    \maybeLabel{#1}{#1/2}{black}
    \maybeLabel{#2}{#1 + #2/2}{black}
    \maybeLabel{#3}{#1 + #2 + #3/2}{black}
\fi

    \addplot [fill=\colA, draw=black] coordinates {(#1,0)};
    \addplot [fill=\colB, draw=black] coordinates {(#2,0)};
    \addplot [fill=\colC, draw=black] coordinates {(#3,0)};

    \end{axis}
\end{tikzpicture}%
}

\newcommand{\rectbar}[3][0.2cm]{%
  \begin{tikzpicture}[baseline=(current bounding box.south)]
    \path[draw=black, line width=0.4pt, fill=#3]
      (0,0) rectangle (#2,#1);
  \end{tikzpicture}%
}

\newcommand{\splitrectbar}[4][0.2cm]{%
  \begin{tikzpicture}[baseline=(current bounding box.south)]
    \path[draw=black, line width=0.4pt, fill=#3]
      (0,0) rectangle (#2/2, #1);
    \path[draw=black, line width=0.4pt, fill=#4]
      (#2/2,0) rectangle (#2, #1);
  \end{tikzpicture}%
}

\def\BibTeX{{\rm B\kern-.05em{\sc i\kern-.025em b}\kern-.08em
    T\kern-.1667em\lower.7ex\hbox{E}\kern-.125emX}}
\AtBeginDocument{\definecolor{ojcolor}{cmyk}{0.93,0.59,0.15,0.02}}
\def\OJlogo{\vspace{-4pt}\hskip-4pt\includegraphics[height=18pt]{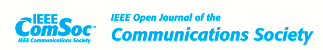}}

\begin{document}

\receiveddate{XX Month, XXXX}
\reviseddate{XX Month, XXXX}
\accepteddate{XX Month, XXXX}
\publisheddate{XX Month, XXXX}
\currentdate{30 January, 2026}
\doiinfo{XXX.XXXX.XXXXX}

\title{Assessing the Real-World Impact of Post-Quantum Cryptography on WPA-Enterprise Networks}

\author{{Lukas Köder\IEEEauthorrefmark{1}\orchid{0000-0003-3620-2336}}, Nils Lohmiller\IEEEauthorrefmark{1}\orchid{0009-0008-1301-4002}, Phil Schmieder\IEEEauthorrefmark{2}\orchid{0009-0007-7196-2791}, Bastian Buck\IEEEauthorrefmark{1}\orchid{0009-0009-9681-727X}, Michael Menth\IEEEauthorrefmark{3}\orchid{0009-0008-1301-4002} \IEEEmembership{(Senior Member, IEEE)}, AND Tobias Heer\IEEEauthorrefmark{1,4}\orchid{0000-0003-3119-252X}}

\affil{University of Applied Sciences Esslingen, Germany}
\affil{University of Wuppertal, Germany}
\affil{University of Tübingen, Germany}
\affil{Belden Inc., Neckartenzlingen, Germany}
\corresp{CORRESPONDING AUTHOR: Lukas Köder (e-mail: lukas.koeder@hs-esslingen.de).}

\authornote{This work has been funded by the Deutsche Forschungsgemeinschaft (DFG, German Research Foundation) – Project-ID 528745080 - FIP~68. 
The authors alone are responsible for the content of the paper.}
\markboth{Preparation of Papers for IEEE OPEN JOURNALS}{Author \textit{et al.}}

\begin{abstract}
The advent of large-scale quantum computers poses a significant threat to contemporary network security protocols, including Wi-Fi Protected Access (WPA)-Enterprise authentication.
To mitigate this threat, the adoption of Post-Quantum Cryptography (PQC) is critical. 
In this work, we investigate the performance impact of PQC algorithms on WPA-Enterprise–based authentication. 
To this end, we conduct an experimental evaluation of authentication latency using a testbed built with the open-source tools FreeRADIUS and hostapd, measuring the time spent at the client, access point, and RADIUS server. 
We evaluate multiple combinations of PQC algorithms and analyze their performance overhead in comparison to currently deployed cryptographic schemes.
Beyond performance, we assess the security implications of these algorithm choices by relating authentication mechanisms to the quantum effort required for their exploitation. 
This perspective enables a systematic categorization of PQ-relevant weaknesses in WPA-Enterprise according to their practical urgency. 
The evaluation results show that, although PQC introduces additional authentication latency, combinations such as ML-DSA-65 and Falcon-1024 used in conjunction with ML-KEM provide a favorable trade-off between security and performance. 
Furthermore, we demonstrate that the resulting overhead can be effectively mitigated through session resumption.
Overall, this work presents a first real-world performance evaluation of PQC-enabled WPA-Enterprise authentication and demonstrates its practical feasibility for enterprise Wi-Fi deployments.
\end{abstract}


\maketitle

\section{INTRODUCTION}
\glsresetall
\IEEEPARstart{W}{ireless} networks are ubiquitous and an essential part of modern society.
Among them, IEEE\,802.11 Wireless Local Area Networks (WLANs), also known as Wi-Fi, are one of the most common types of wireless networks.
To secure these networks, companies utilize WPA-Enterprise (Wi-Fi Protected Access–Enterprise) in conjunction with a) \gls{EAP-TLS}~\cite{eap-tls-rfc5216} or b) \gls{EAP-TTLS}~\cite{rfc5281}, which provide strong, certificate-based authentication using asymmetric cryptography, e.g., RSA or Diffie-Hellman.

\begin{figure}[]
    \centering
    \includegraphics[width=\linewidth]{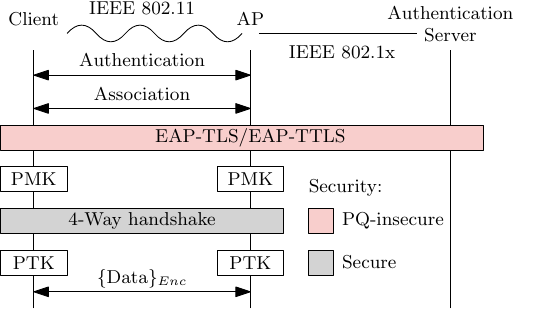}
    \caption{Post-Quantum Security of current EAP-TLS or EAP-TTLS based WPA-Enterprise Authentication \gls{PMK} and \gls{PTK}.}
    \label{fig:wpa-enterprise}
\end{figure}

Unfortunately, the underlying hardness assumptions that make these technologies secure, i.e., integer factorization for RSA and the discrete logarithm problem for Diffie-Hellman, are broken by Shor's quantum factoring algorithm~\cite{shorPolynomialTimeAlgorithmsPrime1997}.
Therefore, these schemes cannot be considered secure in the presence of an adversary with a cryptographically relevant quantum computer.
As a result, the security of WPA-Enterprise networks is threatened by the advent of large scale quantum-computers due to the reliance on asymmetric cryptography in the current \gls{EAP-TLS} and \gls{EAP-TTLS} exchange.
In contrast, the WPA 4-Way handshake, which is based solely on symmetric primitives, remains secure even in a \gls{PQ} setting (cf. Figure~\ref{fig:wpa-enterprise}).

Although large-scale quantum computers do not exist yet, they are no longer perceived as unattainable and are expected to become practical in approximately 15 years (at the time of writing) \cite{BSI2025Quantencomputer}.
However, the long standardization and transition processes of technology threatened by quantum computers, and the threat of \textit{harvest-now-decrypt-later} attacks, call for proactive changes. 

The use of \gls{PQC} algorithms as drop-in replacements for current schemes is a promising solution to ensure continuous, lasting security.
However, most \gls{PQC} algorithms are significantly more demanding regarding their performance than their classical counterparts in terms of memory requirements and computational overhead for key operations like key generation, encapsulation, decapsulation, signing and verification, and thus may not serve as simple replacements regarding the performance.
In order to identify suitable \gls{PQC} algorithms, the \gls{NIST} determines and standardizes \gls{PQ} \glspl{KEM} and signature schemes as part of its \gls{PQC} standardization project.

The exact impact of these algorithms on the WPA-Enterprise authentication duration remains unclear, potentially resulting in additional overhead. 
We address this uncertainty by making the following key contributions:

\begin{itemize}
    \item We introduce and categorize the \emph{quantum annoyance} for different attacks on WPA-Enterprise protected networks. This resilience metric characterizes how the effort required by a quantum adversary scales with the number of quantum computing operations needed to extract security relevant information.
    \item We evaluate the performance impact of \gls{PQC} schemes on WPA-Enterprise authentication based on FreeRADIUS and hostapd under realistic network conditions.
    \item We release the required software changes to FreeRADIUS as well as hostapd and the testbed setup and configuration files on GitHub to promote further research into \gls{PQC}-secure wireless authentication. 
\end{itemize}
The remainder of the paper is structured as follows:
Section~\ref{sec:background} highlights the necessary background regarding the \gls{PQC} algorithms and their impact on the EAP-TLS and EAP-TTLS exchange.
Section~\ref{sec:quantumattacks} categorizes and assesses the impact of quantum attacks on EAP-TLS.
We discuss the experimental setup that we use to measure the feasibility of \gls{PQC} algorithms in WPA-Enterprise networks in Section~\ref{sec:setup}.
Section~\ref{sec:eval} presents the evaluation results, which we discuss in Section~\ref{sec:disc}.
We present related work in Section~\ref{sec:related}.
Finally, Section~\ref{sec:conclusion} concludes the paper.

\section{Background}
\label{sec:background}
The primary countermeasure against quantum attacks is to replace vulnerable algorithms with \gls{PQ}-secure alternatives. 
In TLS, and therefore EAP-TLS/EAP-TTLS, this involves integrating \gls{PQ}-secure key encapsulation and signature schemes into the handshake. 
To assess the authentication and security impact of \gls{PQC} on enterprise wireless networks, we first outline the schemes selected in the \gls{NIST} standardization process.
Afterward, we explain EAP-TLS based authentication and the necessary changes that make it \gls{PQ}-secure.

\subsection{Post-Quantum Secure Schemes}
\begin{table}[]
    \centering
    \caption{The \gls{NIST} \gls{PQ} security levels from \cite{nistpqcp-cfp}.}
    \begin{tabular}{cl}
        \toprule
        Security level & Minimal comparable resource requirements\\
        \midrule
        1 & key search on a block cipher with a 128-bit key \\
        2 & collision search on a 256-bit hash function \\
        3 & key search on a block cipher with a 192-bit key \\
        4 & collision search on a 384-bit hash function  \\
        5 & key search on a block cipher with a 256-bit key \\
        \bottomrule
    \end{tabular}
    \label{tab:nistlevel}
\end{table}

\begin{table*}[]
    \centering
    \caption{Overview of standardized (or to be standardized) signature schemes. Operation cycle measurements were taken from the \gls{PQ} signature zoo project by PQShield \cite{pqshield_nist_sigs_zoo_2024} and the \gls{NIST} submission packages \cite{NISTPQC-R3:SPHINCS+20}.}
    \label{tab:pqc-overview-sigs}
    \begin{tabular}{@{}lcrrrrrc@{}}
        \toprule
        \multirow{2}{*}{Algorithm} & \multirow{2}{*}{Problem type} & \multicolumn{3}{c}{Size (Bytes)} & \multicolumn{2}{c}{Operations (Cycles)} & \multirow{2}{*}{PQ security \cite{nist-pqc}} \\
        \cmidrule(lr){3-5}\cmidrule(lr){6-7} &  & Public key & Secret key & Signature & Sign & Verify &\\
        \midrule
        RSA-2048 & Integer factorization & $\gtrsim 256$ & $\gtrsim 256$ & 256 & 27\,000\,000 & 45\,000 & $-$ \\
        \midrule
        Falcon-512 & \multirow{2}{*}{NTRU SIS} & 897 & 1\,281 & 666 & 1\,009\,764 & 81\,036 & Level 1\\
        Falcon-1024 & & 1\,793 & 2\,305 & 1\,280 & 205\,308\,0 & 160\,596 & Level 5 \\
        \midrule
        ML-DSA-44 & \multirow{3}{*}{Module LWE / SIS}& 1\,312 & 2\,560 & 2\,420 & 333\,013 & 118\,412 & Level 2\\
        ML-DSA-65 & & 1\,952 & 4\,032 & 3\,309 & 529\,106 & 179\,424 & Level 3\\
        ML-DSA-87 & & 2\,592 & 4\,896 & 4\,627 & 642\,192 & 279\,936 & Level 5\\
        \midrule
        SLH-DSA-SHA2-128f & \multirow{3}{*}{Hash functions} & 32 & 64 & 17\,088 & 33\,651\,546 & 2\,150\,290 & Level 1\\
        SLH-DSA-SHA2-192f & & 48 & 96 & 35\,664 & 55\,320\,742 & 3\,492\,210 & Level 3\\
        SLH-DSA-SHA2-256f & & 64 & 128 & 49\,856 & 109\,104\,452 & 3\,559\,052 & Level 5\\
        \midrule
        SLH-DSA-SHA2-128s & \multirow{3}{*}{Hash functions} & 32 & 64 & 7\,856 & 644\,740\,090 & 861\,478 & Level 1\\
        SLH-DSA-SHA2-192s & & 48 & 96 & 16\,224 & 1\,246\,378\,060 & 1\,444\,030 & Level 3\\
        SLH-DSA-SHA2-256s & & 64 & 128 & 29\,792 & 1\,025\,721\,040 & 1\,986\,974 & Level 5\\
        \bottomrule
    \end{tabular}
\end{table*}

\begin{table*}[]
    \centering
    \caption{Overview of standardized key exchange schemes. The cycle count for ML-KEM are taken from~\cite{NISTPQC-R3:CRYSTALS-KYBER20}.}
    \label{tab:pqc-overview-kex}
    \begin{tabular}{@{}lcrrrrrrc@{}}
        \toprule
        \multirow{2}{*}{Algorithm} & \multirow{2}{*}{Problem type} & \multicolumn{3}{c}{Size (Bytes)} & \multicolumn{3}{c}{Operations (Cycles)} & \multirow{2}{*}{PQ security \cite{nist-pqc}} \\
         \cmidrule(lr){3-5}\cmidrule(lr){6-8} & & Public key & Secret key & Ciphertext & KeyGen & Encaps & Decaps &\\
        \midrule
        ECDHKE with X25519 & Discrete logarithm & 64 & 32 & & & & & $-$\\
        \midrule
        ML-KEM-512 & \multirow{3}{*}{Module LWE} & 800 & 1\,632 & 768 & 122\,684 & 154\,524 & 187\,960 & Level 1\\
        ML-KEM-768 & & 1\,184 & 2\,400 & 1\,088 & 199\,408 & 235\,260 & 274\,900 & Level 3\\
        ML-KEM-1024 & & 1\,568 & 3\,168 & 1\,568 & 307\,148 & 346\,648 & 396\,584 & Level 5\\
        \bottomrule
    \end{tabular}
\end{table*}


In their \gls{PQC} project, the \gls{NIST} studies and standardizes signature schemes and \glspl{KEM} believed to be resistant to quantum attacks.  
Specific parameterizations of these studied schemes are assigned \emph{security levels} ranging from level 1 (lowest) to level 5 (highest).
\autoref{tab:nistlevel} shows and classifies the five levels with increasing security expectations as defined by the \gls{NIST} for their \gls{PQC} project \cite{nistpqcp-cfp}.

Since May 2025, the two \gls{KEM}s CRYSTALS-Kyber and HQC and the three signature schemes CRYSTALS-Dilithium, SPHINCS+ and Falcon are officially standardized.
CRYSTALS-Kyber was standardized as ML-KEM in \cite{nist2024fips203} and the signature algorithms CRYSTALS-Dilithium and SPHINCS+ were standardized as ML-DSA in \cite{nist2024fips204} and SLH-DSA in \cite{nist2024fips205}, respectively.
All standardized schemes are available in parameterizations corresponding to the security levels 1/2, 3 and 5.
These schemes were primarily selected because of the high confidence in their security and their well-studied mathematical bases, and not because of their performance.
Consequently, the cost of these algorithms, both in operation speed and artifact size, e.g. signatures or public keys, varies significantly among them, and is generally higher than for classical schemes, e.g., RSA or Diffie-Hellman.
In the following, we discuss the \gls{PQ} algorithms that we evaluate in this paper.
Table~\ref{tab:pqc-overview-sigs} and Table~\ref{tab:pqc-overview-kex} provide an overview of the standardized \gls{PQC} signature schemes and \glspl{KEM} compared to current classical algorithms with respect to the size of their artifacts. 
The following sections outline each of these \gls{PQC} algorithms.

\paragraph{ML-KEM}
ML-KEM was the first \gls{PQ} \gls{KEM} selected by the \gls{NIST} and is currently the only one that has already been standardized.
The security of ML-KEM is based on the presumed hardness of the \gls{LWE} problem over module lattices, also known as \emph{Module-LWE}.
At its core, the \gls{LWE} problem asks to find the mapping of a matrix into a vector space given a noisy sample of said mapping~\cite{EPRINT:Lyubashevsky24}.
This problem is presumed to be hard even for quantum computers.

The artifact sizes of ML-KEM are substantially larger than those of a contemporary Diffie–Hellman key exchange based on the X25519 curve. 
In particular, the public key size increases by 1150\,\% from 64\,bytes to 800\,bytes for security level~1, by 1750\,\% to 1184\,bytes for security level~3, and by 2350\,\% to 1568\,bytes for security level~5, respectively (cf. Table~\ref{tab:pqc-overview-kex}). 
These increases are especially problematic in deployment scenarios with low available data rates.

\paragraph{ML-DSA}
ML-DSA is the primary signature scheme selected by the \gls{NIST} in their \gls{PQC} standardization process \cite{NISTIR8413-upd1}.
ML-DSA is a lattice-based digital signature scheme using the Fiat-Shamir paradigm with aborts.
The security of ML-DSA is based on the presumed hardness of the \emph{Module-LWE} problem and the \emph{Module \gls{SIS}} problem.
The \gls{SIS} problem involves finding short integer solutions to systems of linear equations under a modulus \cite{EPRINT:Lyubashevsky24} and is presumed to be computationally hard even against quantum adversaries.

The \gls{NIST} selected ML-DSA for standardization because of its high efficiency, easy implementation, and well-studied theoretical security basis.
Compared to the other signature schemes selected for standardization, ML-DSA performs well in both speed and size requirements (see \autoref{tab:pqc-overview-sigs}).
Therefore, ML-DSA should be considered as a valid candidate for a broad range of applications.

\paragraph{SLH-DSA}
The second standardized \gls{PQ} signature scheme is SLH-DSA, a stateless hash-based signature scheme.
Unlike ML-DSA and Falcon, the security of SLH-DSA is not based on the assumed quantum hardness of lattice problems, but on the security of hash functions.
Although attacks on hash functions using Grover’s quantum search algorithm~\cite{groverQuantumAlgorithm} are significantly more efficient than classical attacks, their impact can be mitigated by increasing the security level of the affected hash functions.
This is because Grover’s algorithm provides only a quadratic speedup for brute-force search, so doubling the hash output length suffices to restore the original level of security against quantum adversaries.

SLH-DSA was standardized with 12 different parameter sets yielding different security levels, artifact sizes, and performance.
Each instantiation for the \gls{NIST} security levels 1, 3, and 5 exist in a \emph{small} (suffix \textbf{s}) and a \emph{fast} (suffix \textbf{f}) variant trading computational speed for artifact size respectively.
For example: SLH-DSA-*-128s has a much smaller signature size than SLH-DSA-*-128f, but its CPU performance is much slower~\cite{cloudflare-blog}.
Additionally, the parameter sets for SLH-DSA are standardized for both the SHA-2 and SHAKE hash functions, resulting in 12 different variations.
In general, SLH-DSA has a larger signature size and a smaller public key size, compared to Falcon and ML-DSA.
\autoref{tab:pqc-overview-sigs} shows only SLH-DSA instantiations using SHA-2, since it is the default hash algorithm for SLH-DSA in the OpenSSL implementation employed in this work.

\paragraph{Falcon}
The Falcon signature scheme is the third \gls{PQ} signature algorithm selected for standardization.
Unlike ML-DSA and SLH-DSA, Falcon is not yet standardized.
Falcon is a lattice-based scheme that utilizes the \textit{hash-and-sign} paradigm.
Its security is based on the presumed hardness of the \gls{SIS} problem over specific, so-called \emph{NTRU} lattices.
Of all selected signature schemes, Falcon yields the smallest public keys and signatures.
Additionally, Falcon verifies signatures faster than SLH-DSA and ML-DSA.
Therefore, Falcon may be the superior choice in heavily bandwidth-constrained applications.
Falcon is considerably harder to securely implement than other signature schemes like ML-DSA.
This is both because of the use of complex data structures used in Falcon and because of the use of floating-point arithmetic.

\paragraph{Hybrid Constructions}
The mathematical foundations and assumptions of many \gls{PQ} schemes like ML-KEM, ML-DSA, or Falcon are relatively novel in contrast to established schemes like RSA or ECDSA that are based on older, thoroughly analyzed foundations.
Therefore, these new schemes, although presumed to be \gls{PQ}-secure, have yet to earn similar levels of trust in their security.
To mitigate this trust gap, cryptographic constructions can be derived that combine an established scheme with a \gls{PQ}-secure scheme.
These constructions are referred to as \emph{hybrid} schemes.
The goal of hybridization is to achieve a scheme that remains secure even if a constituent scheme is broken.
A \gls{PQ}-hybrid scheme provides a risk-aware approach during the transition process to post-quantum cryptography by offering protection against both current and future quantum adversaries and possible advances in breaking any constituent scheme.
Hybrid \glspl{KEM} aim to combine \glspl{KEM} to construct a single \gls{KEM} that remains secure if one of the constituent \glspl{KEM} is broken.
Regardless of the combiner function and the properties of the underlying \glspl{KEM}, the resulting hybrid \gls{KEM} can preserve the security guarantees of its weakest components.
Hybrid signature schemes combine a \gls{PQ} and a classical signature algorithm such that the resulting scheme remains secure even if the \gls{PQ} signature is broken or the classical signature is compromised by a quantum computer.
Both the hybrid \glspl{KEM} and the signature schemes we investigate in this paper are constructed using a simple concatenation combiner, where the \gls{PQ} shared secret (or signature) is appended to its classical counterpart.

\subsection{Post Quantum EAP-TLS}
\label{sec:eap-tls}

EAP-TLS is an authentication method based on mutual certificate authentication, using the TLS handshake to negotiate cryptographic algorithms and derive shared keying material. 
The integration of \gls{PQC} into EAP-TLS mandates TLS 1.3 which introduces changes to the handshake structure, cipher suites, and key schedule~\cite{eap-tls-1-3-rfc9190}~\cite{tls-1-3-rfc8446}.
For a \gls{PQ}-secure EAP-TLS or EAP-TTLS exchange, the cryptographic primitives such as RSA or DH are replaced with \gls{PQ}-safe alternatives to protect against adversaries with quantum computers.

\begin{figure}[]
    \centering
    \includegraphics[width=\linewidth]{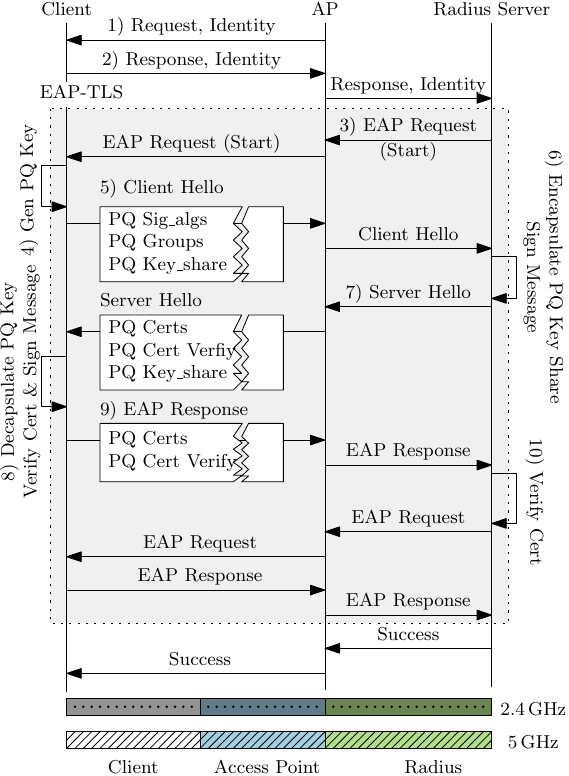}
    \caption{\gls{PQ}-EAP-TLS message exchange between a client and a RADIUS server over Wi-Fi. Large messages must be fragmented into several \gls{EAP} messages. The colors shown will be used consistently for all measurements in the 2.4 GHz and 5 GHz bands throughout this work.}
    \label{fig:eap-tls}
    \vspace{-1.5em}
\end{figure}

Figure~\ref{fig:eap-tls} outlines the message flow of a \gls{PQ}-secure EAP-TLS exchange in a Wi-Fi environment, involving a client~ \splitrectbar{0.5cm}{client5}{client}, an \gls{AP}~\splitrectbar{0.5cm}{ap5}{ap}, and a RADIUS server~\splitrectbar{0.5cm}{radius5}{radius}. 
The RADIUS protocol, which typically runs over UDP, lacks guaranteed message delivery; thus, every EAP-message must be acknowledged with an EAP-Response to signal reliable transmission.
These two messages are referred to as an \gls{EAP} round-trip.
After successful association, the AP initiates authentication by sending an EAP-Request/Identity to the client. 
The client responds with an EAP-Response/Identity, which the \gls{AP} relays to the RADIUS server. 

If the identity is accepted, the server begins the EAP-TLS handshake by sending an EAP-Request containing the TLS \textit{ClientHello} message. 
This message includes \gls{PQ} signature algorithm identifiers, e.g., ML-DSA, SLH-DSA, Falcon, supported \gls{PQ} key exchange groups, e.g., ML-KEM-512, ML-KEM-768, ML-KEM-1024, and the corresponding key share of the client.
Due to the increased size of \gls{PQC} artifacts, fragmentation of handshake messages might be necessary, resulting in additional \gls{EAP} round-trips and therefore latency. 
The server responds with a \textit{ServerHello} encapsulated inside an EAP-Request Message, which contains the \gls{PQ} certificate chain of the server (\gls{PQ} Certs), its \gls{PQ} signature (\gls{PQ} Cert Verify), and the corresponding key share ciphertext (\gls{PQ} Key\_share). 
Here, the larger certificate sizes may also lead to fragmentation and overhead during transmission.
Upon reception, the client performs a decapsulation operation using the received \gls{PQ} ciphertext and verifies the identity of the server using the certificate. 
If successful, the client responds with its own \gls{PQ} certificate chain in an EAP-Response. 
The server verifies the credentials of the client and upon success proceeds with the standard IEEE 802.11i 4-way Handshake to derive the \gls{PTK}.

Overall, a \gls{PQ}-secure EAP-TLS exchange introduces latency across three main stages: a) client-side processing, which includes \gls{PQ} cryptographic operations as well as modulation, demodulation, retransmissions, and MAC-layer contention b) \gls{AP} forwarding and local processing overhead, similar to the client-side, but without the cryptographic operations and c) RADIUS server-side processing, which involves cryptographic computations and identity verification.

\section{Quantum Attacks on EAP-TLS}
\label{sec:quantumattacks}
An adversary with access to a cryptographically relevant quantum computer threatens the EAP-TLS message exchange in three ways. 
The attacker can: a) break the security of the client certificate, which would allow the attacker to authenticate and join the network b) break the security of the server certificate, which would enable \gls{MITM} attacks (the attacker intercepts and relays communications between server and client) or “evil twin” attacks (the attacker impersonates a legitimate Wi-Fi network to trick clients into connecting to it) and c) break the security of the key exchange, which would allow the attacker to derive the \gls{PTK} and thereby decrypt or spoof traffic. 
Each of these attacks has different implications, varying in both severity and duration of impact.
For instance, compromising a key share or session key affects the confidentiality and integrity of only a single session, since these keys are refreshed with each new authentication. 
In contrast, compromising a client or server certificate has a long-lasting impact, as certificates are typically not updated frequently.
If a client certificate is compromised, the integrity of client communications will be affected.
Similarly, a compromised server certificate will affect the confidentiality, integrity, and availability of server communications.
As a result, legitimate servers may become unavailable, clients may be redirected to rogue networks, and the confidentiality and authenticity of previously trusted connections are undermined by the attacker.

We build upon the quantum annoying property introduced by Thomas~\cite{thomasannoying} and Eaton et al.~\cite{eaton2021quantum}.
We further develop it by categorizing attacks according to their quantum annoyance level, ranging from low (bad) to high (good).
The term \textit{quantum annoyance}, describes the significant obstacle that even an idealized quantum adversary (with access to Shor's or Grover's algorithm) must overcome to extract security-relevant information from a complex communication structure in a targeted manner.
In simple terms, the quantum annoyance factor increases with the number of times a quantum computer must solve the discrete logarithm problem in order to extract meaningful information during an attack.

Table~\ref{tab:quantum-annoyance} categorizes the outlined \gls{PQ}-threats based on their impact, and we qualitatively assign them different quantum annoyance values, ranging from low (quantum computer needs few operations to attack valuable information, which is undesirable) to high (quantum computer needs many operations to attack valuable information, making them more resistant to attacks). 

\begin{table}[]
\centering
\caption{Post quantum attack targets and their quantum annoyance and the impact on the confidentiality (C), integrity (I), and availability (A).}
\label{tab:quantum-annoyance}
\begin{tabular}{@{}lclllll@{}}
\toprule
\multirow{2}{*}{Post Quantum attack target} & \multirow{2}{*}{Quantum annoyance} & \multicolumn{3}{c}{Impact}                                                          \\ \cmidrule(lr){3-5} 
        &                                    & \textbf{C} & \textbf{I} & \multicolumn{1}{l}{\textbf{A}}  \\ \midrule
Client certificate          & Medium                             & --         & X         & --                                \\
Server certificate          & Low                                &  X          & X           &   X               \\
Key share/Session key                   & High                               & X          & X          &   --              \\ \bottomrule 
\end{tabular}
\end{table}

We classify an attack against the client certificate as having a medium level of quantum annoyance, since the attacker must solve the underlying hard problem for each client certificate. 
With a compromised certificate, an attacker can impersonate one client; to impersonate additional clients, further certificates must be attacked. 
The server certificate, by contrast, is classified as having a low level of quantum annoyance, because a single corrupted server certificate is sufficient to launch an evil twin attack. 
An attack against the key share or session key has a high level of quantum annoyance, as the attacker must solve the underlying hard problem for each key in order to decrypt data until the next authentication occurs.

Categorizing and defining the quantum annoyance for different threats underscores the importance of integrating \gls{PQ}-secure algorithms into EAP-TLS/EAP-TTLS and, consequently, WPA-Enterprise. 
In particular, the low quantum annoyance of server certificates shows that with comparatively little quantum effort, significant damage can be inflicted such as \gls{MITM} attacks that compromise the confidentiality, integrity, and availability of users. 
Therefore, transitioning to \gls{PQ}-secure algorithms is crucial. 
However, since \gls{PQC} algorithms vary considerably in artifact size and performance, their suitability for WPA-Enterprise authentication must be carefully assessed. 
The following sections evaluate the integration of different \gls{PQC} schemes into EAP-TLS/EAP-TTLS and examine their impact on both the performance and deployment of WPA-Enterprise authentication.

\section{Experimental Setup}
\label{sec:setup}

This section outlines the experimental setup we use to show the feasibility and influence of \gls{PQC} on WPA-Enterprise authentication.
The testbed is designed to resemble a small enterprise network consisting of a supplicant (the client), an authenticator (the AP), and an authentication server (the RADIUS server).
We perform the measurements using two Protectli mini computers (one AP and one RADIUS Server) and one consumer laptop (the client). 
The client runs Ubuntu 25.04 kernel version 6.14.0-15-generic and uses the linux userspace program wpa\_supplicant version 2.12 to connect and authenticate to the Wi-Fi network. 
The RADIUS server is based on the open-source RADIUS implementation FreeRADIUS version 3.2.8 and runs Ubuntu 25.04 kernel version 6.14.0-22-generic. 
Both Protectli mini computers are equipped with an Intel Celeron J6412, 8\,GB RAM, with the access point using a MediaTek MT7921AUN wireless chipset.
The client features an AMD Ryzen 5 5500U, 16\,GB RAM, and Realtek RTL8821CE wireless network adapter.
All devices inside the testbed run OpenSSL version 3.4.1, use liboqs version 0.14.0, and oqs-provider version 0.9.0.
Note that the larger cryptographic artifacts of the \gls{PQC} algorithms require small changes to the hostapd and FreeRADIUS source code.
These changes include increasing the allowed number of \gls{EAP} round trips from 50 to 500 and removing TLS length limitation to facilitate the significantly larger certificate sizes of the \gls{PQC} algorithms. 
We provide these patched versions, the test scripts, and exemplary pcap files of this work on GitHub\,\footnote{\url{https://github.com/hs-esslingen-it-security/wpa-enterprise-pqc}}, allowing other researchers to easily set up their own WPA-Enterprise testbed.

The experimental setup operates in both the 2.4\,GHz and 5\,GHz bands with a fixed 20\,MHz channel width, to reduce the chance of interference and improve link stability. 
Measurements are performed on channel 1 for the 2.4\,GHz band and channel 36 for the 5\,GHz band, both commonly used in indoor deployments.
Minimum data rates are configured at the standard 1\,Mbps at 2.4\,GHz and 6\,Mbps at 5\,GHz, to facilitate connectivity even under degraded conditions.

To evaluate the performance of WPA-Enterprise authentication under realistic network conditions, we conduct experiments across three wireless signal quality situations. 
These situations cover typical link conditions in real-world deployments, from strong and stable connections to weak and unreliable ones.

\begin{figure}[]
    \centering
    \includegraphics[width=\linewidth]{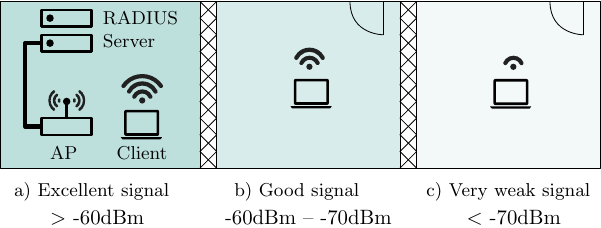}
    \caption{Testbed setup for three evaluation situations: a) Excellent signal ($>$ –60\,dBm) near the access point, b) good signal (–60\,dBm to –70\,dBm) in the adjacent room, and c) very weak signal ($<$ –70\,dBm).}
    \label{fig:testbed}
\end{figure}

We adopt the signal classification of Kahn et al.~\cite{khan2024wifi} resulting in the following evaluation situations (cf.~Figure~\ref{fig:testbed}):
\begin{enumerate}
    \item[a)] \emph{Excellent Signal} The client and \gls{AP} are positioned in the same room, with minimal physical separation. The signal exceeds –60\,dBm, corresponding to a stable, high-quality connection. 
\item[b)] \emph{Good Signal} The client is located at a greater distance from the \gls{AP}.
This spatial separation lowers the signal quality to between –60\,dBm and –70\,dBm, representing a moderate, but still reliable, connection.
\item[c)] \emph{Very Weak Signal} The client and \gls{AP} are positioned even farther apart than before, separated by at least one wall.
This additional attenuation reduces the signal to below –70\,dBm.
\end{enumerate}

We conduct all experiments in a realistic indoor office environment on a university campus where multiple external Wi-Fi networks are simultaneously active. 
These uncontrolled signals introduce background interference and contention for the wireless medium, resulting in  practical, real-world conditions rather than isolated laboratory settings.

\begin{table*}[]
\setlength{\tabcolsep}{2pt}
\centering
\caption{Authentication component breakdown across algorithms (\rectbar{0.5cm}{24ghz}~2.4 GHz vs. \rectbar{0.5cm}{5ghz}~5 GHz): \splitrectbar{0.5cm}{ap5}{ap}~Access Point, \splitrectbar{0.5cm}{radius5}{radius}~Radius Server, \splitrectbar{0.5cm}{client5}{client}~Client.}
\begin{tabularx}{\textwidth}{lccXccXccX}
\toprule
&\multicolumn{3}{c}{EAP time (ms) situation a)} &\multicolumn{3}{c}{EAP time (ms) situation b)} &\multicolumn{3}{c}{EAP time (ms) situation c)}\\\cmidrule(lr){2-4}\cmidrule(lr){5-7}\cmidrule(lr){8-10}
Algorithm &50$^{th}$&  95$^{th}$ & Time composition &50$^{th}$& 95$^{th}$ & Time composition &50$^{th}$& 95$^{th}$ & Time composition\\
\midrule

RSA-2048
& 0.11 & 0.12 & \stackedbar{0.02429997920989985}{0.0415616035461425}{0.039278388023376395}{1}
& 0.14 & 0.18 & \stackedbar{0.045042872428894}{0.04147088527679435}{0.05094230175018305}{1}
& 0.17 & 0.22 & \stackedbar{0.07902419567108149}{0.0405534505844116}{0.0497697591781616}{1} \\

& 0.08 & 0.1 & \stackedbar{0.023574233055114698}{0.04749643802642815}{0.00959980487823485}{0}
& 0.08 & 0.11 & \stackedbar{0.0236512422561645}{0.0441608428955078}{0.00971353054046625}{0}
& 0.09 & 0.1 & \stackedbar{0.03874039649963375}{0.04338753223419185}{0.010101318359374951}{0} \\

\midrule
\multicolumn{10}{l}{\textbf{Security level 1/2}}\\

Falcon-512
& 0.2 & 0.21 & \stackedbar{0.0310761928558349}{0.08908855915069575}{0.07768392562866211}{1}
& 0.24 & 0.27 & \stackedbar{0.0656551122665405}{0.08904373645782465}{0.08417141437530515}{1}
& 0.35 & 0.43 & \stackedbar{0.15477764606475825}{0.08526730537414545}{0.1093488931655883}{1} \\

& 0.13 & 0.14 & \stackedbar{0.029568552970886203}{0.08450591564178464}{0.0166629552841186}{0}
& 0.13 & 0.14 & \stackedbar{0.0295101404190063}{0.0873996019363403}{0.0161638259887695}{0}
& 0.16 & 0.17 & \stackedbar{0.0546374320983886}{0.08985090255737299}{0.016520142555236747}{0} \\

ML-DSA-44
& 0.27 & 0.3 & \stackedbar{0.0410506725311279}{0.08527147769927976}{0.14568948745727534}{1}
& 0.37 & 0.44 & \stackedbar{0.10234808921813959}{0.0893816947937011}{0.1755263805389404}{1}
& 0.57 & 0.69 & \stackedbar{0.29431402683258057}{0.0850480794906616}{0.19007682800292963}{1} \\

& 0.15 & 0.17 & \stackedbar{0.03934717178344725}{0.0837324857711792}{0.0306407213211059}{0}
& 0.15 & 0.16 & \stackedbar{0.03951656818389885}{0.0866638422012329}{0.0285086631774902}{0}
& 0.22 & 3.19 & \stackedbar{0.10204589366912835}{0.08957874774932856}{0.030575633049011203}{0} \\

SLH-DSA-128f
& 1.04 & 1.08 & \stackedbar{0.12390422821044915}{0.29850995540618896}{0.6090528964996338}{1}
& 1.37 & 1.46 & \stackedbar{0.378373384475708}{0.3324931859970093}{0.6647893190383911}{1}
& 2.39 & 2.57 & \stackedbar{1.2437162399291992}{0.3076937198638916}{0.8227782249450684}{1} \\

& 0.53 & 0.55 & \stackedbar{0.11212754249572751}{0.3039342164993286}{0.1155323982238769}{0}
& 0.54 & 0.56 & \stackedbar{0.1143226623535156}{0.3184816837310791}{0.11247944831848145}{0}
& 0.82 & 0.86 & \stackedbar{0.3767356872558594}{0.32605719566345215}{0.11397731304168701}{0} \\

SLH-DSA-128s
& 1.4 & 1.44 & \stackedbar{0.23341763019561765}{0.8640708923339844}{0.3057217597961426}{1}
& 1.89 & 1.98 & \stackedbar{0.6819862127304077}{0.8798935413360596}{0.3320220708847046}{1}
& 1.95 & 2.08 & \stackedbar{0.7056699991226196}{0.8643133640289307}{0.38134264945983887}{1} \\

& 1.14 & 1.16 & \stackedbar{0.22906112670898435}{0.8449807167053223}{0.0651357173919677}{0}
& 1.15 & 1.18 & \stackedbar{0.2277480363845825}{0.864234209060669}{0.0629782676696777}{0}
& 1.29 & 3.48 & \stackedbar{0.35156285762786865}{0.873532772064209}{0.06077551841735835}{0} \\

\midrule
\multicolumn{10}{l}{\textbf{Security level 3}}\\

ML-DSA-65
& 0.34 & 0.36 & \stackedbar{0.0494873523712158}{0.0995508432388305}{0.19341707229614255}{1}
& 0.44 & 0.52 & \stackedbar{0.12540423870086664}{0.1061060428619384}{0.2073097229003906}{1}
& 0.74 & 0.83 & \stackedbar{0.38342154026031494}{0.1038039922714233}{0.2538756132125854}{1} \\

& 0.19 & 0.21 & \stackedbar{0.0445359945297241}{0.1004184484481811}{0.04184401035308835}{0}
& 0.19 & 0.2 & \stackedbar{0.044975519180297796}{0.10269749164581295}{0.037523508071899345}{0}
& 0.28 & 3.24 & \stackedbar{0.12677478790283198}{0.10718929767608634}{0.0396041870117187}{0} \\

SLH-DSA-192f
& 1.99 & 2.07 & \stackedbar{0.22810614109039304}{0.49451422691345215}{1.2512966394424438}{1}
& 2.65 & 2.83 & \stackedbar{0.714120626449585}{0.5788764953613281}{1.3572112321853638}{1}
& 4.64 & 4.97 & \stackedbar{2.4073673486709595}{0.5547267198562622}{1.6952401399612427}{1} \\

& 0.98 & 1.04 & \stackedbar{0.20988905429840085}{0.5148812532424927}{0.25789666175842285}{0}
& 1.0 & 1.03 & \stackedbar{0.21025419235229487}{0.548949122428894}{0.2349550724029541}{0}
& 1.54 & 1.57 & \stackedbar{0.7568552494049072}{0.5538985729217529}{0.22967517375946045}{0} \\

SLH-DSA-192s
& 2.57 & 2.66 & \stackedbar{0.4206323623657226}{1.5703531503677368}{0.5826840400695801}{1}
& 3.3 & 3.44 & \stackedbar{1.0330333709716797}{1.6650809049606323}{0.6147935390472412}{1}
& 3.67 & 3.85 & \stackedbar{1.3380444049835205}{1.6431660652160645}{0.6799079179763794}{1} \\

& 2.14 & 2.17 & \stackedbar{0.4143667221069336}{1.6028026342391968}{0.11969363689422605}{0}
& 2.18 & 2.23 & \stackedbar{0.4136373996734619}{1.6489686965942383}{0.1225323677062988}{0}
& 2.43 & 8.41 & \stackedbar{0.6593799591064453}{1.6426748037338257}{0.11528325080871575}{0} \\

\midrule
\multicolumn{10}{l}{\textbf{Security level 5}}\\

Falcon-1024
& 0.28 & 0.31 & \stackedbar{0.037637710571289}{0.1263042688369751}{0.12292933464050285}{1}
& 0.37 & 0.42 & \stackedbar{0.0921524763107299}{0.1305841207504272}{0.14205658435821528}{1}
& 0.53 & 0.66 & \stackedbar{0.24968194961547846}{0.12576878070831293}{0.16448271274566645}{1} \\

& 0.19 & 0.2 & \stackedbar{0.03696358203887935}{0.1229883432388305}{0.02761840820312495}{0}
& 0.19 & 0.21 & \stackedbar{0.0366520881652832}{0.12872314453124994}{0.02609109878540035}{0}
& 0.24 & 0.25 & \stackedbar{0.08406567573547355}{0.1273648738861084}{0.025855422019958448}{0} \\

ML-DSA-87
& 0.45 & 0.46 & \stackedbar{0.06054127216339105}{0.1252537965774536}{0.25800907611846924}{1}
& 0.58 & 0.63 & \stackedbar{0.1631163358688354}{0.1308110952377319}{0.28482878208160395}{1}
& 0.99 & 1.17 & \stackedbar{0.5254726409912109}{0.1289805173873901}{0.34071290493011475}{1} \\

& 0.24 & 0.25 & \stackedbar{0.05744981765747065}{0.12316048145294184}{0.0517811775207519}{0}
& 0.24 & 0.25 & \stackedbar{0.058340311050415}{0.13036882877349848}{0.049905300140380804}{0}
& 0.34 & 4.91 & \stackedbar{0.1625854969024658}{0.13037014007568354}{0.0511538982391357}{0} \\

SLH-DSA-256f
& 2.68 & 2.94 & \stackedbar{0.31143271923065186}{0.6256681680679321}{1.7446123361587524}{1}
& 3.68 & 3.8 & \stackedbar{1.0601489543914795}{0.8059372901916504}{1.8364553451538084}{1}
& 5.91 & 6.37 & \stackedbar{2.974265694618225}{0.7769674062728882}{2.1165802478790283}{1} \\

& 1.23 & 1.39 & \stackedbar{0.28941214084625244}{0.6197880506515503}{0.33508205413818354}{0}
& 1.42 & 1.47 & \stackedbar{0.2967274188995361}{0.7848765850067139}{0.34121787548065186}{0}
& 2.17 & 2.22 & \stackedbar{1.072513222694397}{0.7786041498184204}{0.32176506519317627}{0} \\

SLH-DSA-256s
& 3.14 & 3.25 & \stackedbar{0.47561812400817866}{1.6379210948944092}{1.0384058952331543}{1}
& 4.13 & 4.28 & \stackedbar{1.3139610290527344}{1.7324851751327515}{1.0938184261322021}{1}
& 4.91 & 5.1 & \stackedbar{2.0180400609970093}{1.6921828985214233}{1.1972708702087402}{1} \\

& 2.28 & 2.33 & \stackedbar{0.4626293182373047}{1.6143800020217896}{0.19667756557464594}{0}
& 2.37 & 2.43 & \stackedbar{0.45230162143707275}{1.7073590755462646}{0.2171852588653564}{0}
& 2.8 & 3.04 & \stackedbar{0.8971036672592163}{1.699237942695618}{0.20367753505706787}{0} \\

\bottomrule
\end{tabularx}
\label{tab:auth-components}
\end{table*}

\section{Evaluation}
\label{sec:eval}
In the following, we present the evaluation results obtained from the testbed described in the previous section. 
The evaluation focuses on the impact of \gls{PQC} on WPA-Enterprise authentication and covers three aspects: 1) the \gls{EAP-TLS} authentication duration, 2) \gls{PQ} \gls{EAP-TLS} session resumption, and 3) hybrid \gls{EAP-TLS} and \gls{EAP-TTLS} configurations.

All measurements are conducted under a consistent experimental setup.
Signature algorithms are evaluated together with \glspl{KEM} that match their respective security levels, i.e., a signature algorithm of level 3 is used with ML-KEM-768 while, a signature algorithm of level 5 is used with ML-KEM-1024. 
The baseline uses RSA-2048 certificates, which are the default in FreeRADIUS and the key exchange is performed using ECDHE with the X25519 curve. 
The client data rate is not fixed, and each authentication experiment is repeated 100 times to account for variability and ensure statistically meaningful results.
We always show median values if not otherwise mentioned, as these values are more robust against statistical outliers.

\begin{figure}[]
    \includegraphics[]{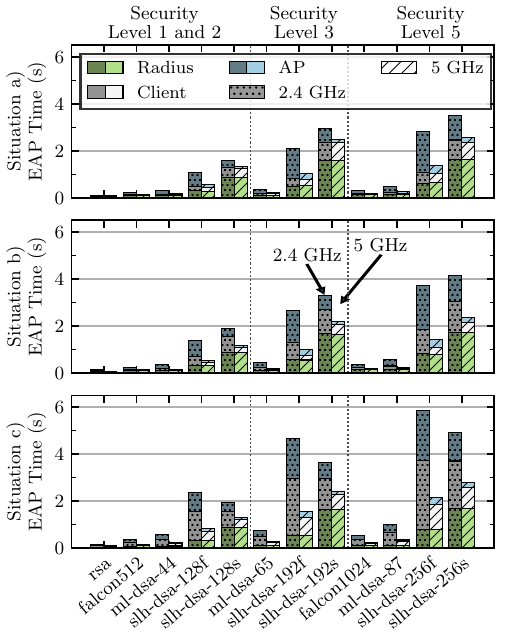}
    \caption{Median EAP-TLS duration of the three evaluation situations. The left bar represents the 2.4\,GHz measurement and the right bar the 5\,GHz.}
    \label{fig:eap-eval-time}
\end{figure}

\subsection{Post Quantum EAP-TLS Authentication Time}
The measured EAP-TLS duration includes the time from the first EAP-Request message until the EAP-Success message.
The EAP-TLS authentication consists of three communication parts that introduce latency namely, 1) the client \splitrectbar{0.5cm}{client5}{client}, 2) the \gls{AP} \splitrectbar{0.5cm}{ap5}{ap}, and 3) the RADIUS server \splitrectbar{0.5cm}{radius5}{radius} (see Section~\ref{sec:background}).

Table~\ref{tab:auth-components} presents the median and 95$^{th}$-percentile duration of an EAP-TLS authentication based on the three outlined evaluation situations. 
The first row of each algorithm always represents the 2.4\,GHz frequency band and the second row the 5\,GHz frequency band.
It is evident that the EAP-TLS authentication is faster across all measurements for the 5\,GHz frequency band.
The faster authentication is caused by the higher standard data rate of 6\,Mbps compared to the lower rate of 1\,Mbps in the 2.4\,GHz frequency band.  
This speedup is more prominent for algorithms with large signatures and higher security levels, such as SLH-DSA-f.

Falcon and ML-DSA exhibit performance closest to RSA-2048, with median \gls{EAP} authentication times of 0.20\,s and 0.27\,s, respectively, for security levels 1/2 in the 2.4\,GHz band, and 0.34\,s (ML-DSA) and 0.28\,s (Falcon) for security levels 3 and 5, respectively.
These times are further reduced in the 5\,GHz band due to the higher minimal datarates resulting in authentication times of 0.13\,s and 0.15\,s for security level 1/2, and 0.19\,s (ML-DSA-66) and 0.19\,s (Falcon1024) respectively.
Note, even though the required authentication time for both algorithms is slower compared to RSA-2048, the observed latencies remain well within acceptable bounds for enterprise deployments.

The RADIUS duration part during the EAP-TLS authentication remains largely unchanged across the two frequency bands, as it is not affected by the data rate of the wireless transmission.
As expected, the computationally demanding algorithms, like SLH-DSA-s, require significantly more time at the RADIUS server compared to the faster alternatives like Falcon, ML-DSA, SLH-DSA-f. 
However, the SLH-DSA-f variants require more messages, potentially being problematic in environments where the RADIUS server is located further away with long round trip times or the wireless transmission is unreliable, resulting in slow authentication times, cf. Table~\ref{tab:auth-components} evaluation situation c). 
Smaller signatures also result in reduced airtime consumption, which can be beneficial in crowded and busy environments.  

The measured \gls{EAP} duration increases under weaker signal conditions, cf. Table~\ref{tab:auth-components} situation c), requiring multiple packet retransmissions due to packet errors and collisions. 
This time increase is visible across all signature algorithms, but exacerbated with increasing security levels and especially prominent for algorithms with large signatures like SLH-DSA-f.
In situation c) and the 2.4\,GHz band, the SLH-DSA-s variant performs better for security level 3 and security level 5 than its computationally faster counterpart.

Figure~\ref{fig:eap-eval-time} puts the evaluation results into perspective, providing an visual overview of the authentication on the same timescale, allowing for a better visual comparison across all evaluation situations.
For each algorithm, the left bar corresponds to the 2.4\,GHz band and the right to the 5\,GHz band.

\textbf{Key Results:} \emph{Smaller signatures are advantageous in weak or unreliable wireless environments, because they require fewer packets. 
Faster SLH-DSA-f variants reduce the computational requirements but require more message exchanges, which can be problematic under weak signal conditions. 
Higher data rates substantially reduce \gls{EAP} authentication times, especially for large signatures like SLH-DSA-f. 
Finally, while slower than RSA, Falcon and ML-DSA show the most favorable results, making them strong candidates for \gls{PQ}-secure replacements.}

\subsection{Post Quantum EAP-TLS Session Resumption}
EAP-TLS session resumption allows the client and authenticator to skip the computationally expensive handshake, but the client still needs to generate and transmit a key share that corresponds to the selected security level.
Because \gls{PQC} schemes use larger key material, we expect this to introduce additional overhead during session resumption.

\begin{figure}
    \includegraphics[]{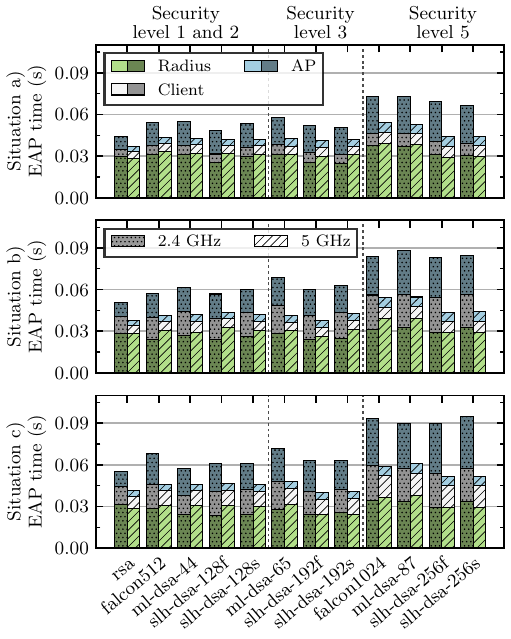}
    \caption{Median EAP-TLS session resumption duration of the three evaluation situation. The left bar represents the 2.4\,GHz measurement and the right bar the 5\,GHz.}
    \label{fig:eap-eval-time-resumption}
\end{figure}

Figure~\ref{fig:eap-eval-time-resumption} shows the evaluation results for the 2.4\,GHz and the 5\,GHz frequency bands across the different evaluation situations.
The measurements across the frequency bands reveal no substantial differences, apart from the expected airtime reduction at 5\,GHz, due to the increased minimum data rate which slightly shortens the overall resumption time.
The relative differences between the three evaluation situations remain consistent with the authentication phase: situation a) outperforms situation b), and situation b) slightly outperforms situation c). 
Resumption time increases with the security level, due to the larger ML-KEM key shares exchanged.
Overall, the evaluation of \gls{PQC} in the context of EAP-TLS session resumption shows that the performance impact is minimal.
While RSA achieves slightly faster resumption times, the difference compared to \gls{PQC} algorithms remains negligible.
Across all tested \gls{PQC} schemes, session resumption times remain similar, with security levels 1 and 3 exhibiting nearly identical performance and only higher security levels showing a marginal slowdown due to the additional overhead from ML-KEM.

\begin{table}[]
\centering
\caption{Storage requirements for each session reported by FreeRADIUS for stateful EAP-TLS session resumption.}
\label{tab:session-cache-size}
\begin{tabular}{lrr}
\toprule
Algorithm    & Storage size (Bytes) & Increase over RSA \\ \midrule
RSA-2048      & 1\,198 &  $-$       \\
\midrule
\textbf{Security level 1+2}& &      \\
Falcon512    & 2\,210&  85\,\%     \\
ML-DSA-44    & 4\,398&  267\,\%    \\
SLH-DSA-128s & 8\,542 & 613\,\%    \\
SLH-DSA-128f & 17\,774 &  1\,384\,\% \\ \midrule
\textbf{Security level 3}& &      \\
ML-DSA-65    & 5\,927&  395\,\%    \\
SLH-DSA-192s & 16\,926 & 1\,313\,\%    \\
SLH-DSA-192f & 36\,366&  2\,935\,\% \\ \midrule
\textbf{Security level 5}& &      \\
Falcon1024    & 3\,720&  211\,\%     \\
ML-DSA-87    & 7\,885&  558\,\%    \\
SLH-DSA-256s & 30\,510 & 2\,447\,\%    \\
SLH-DSA-256f & 50\,574 &  4\,122\,\% \\

\bottomrule
\end{tabular}
\end{table}

EAP-TLS session resumption can be stateless or stateful. 
Stateful session resumptions requires the RADIUS server to store client information for subsequent authentications, while stateless authentication stores the necessary information on the client. 

Table~\ref{tab:session-cache-size} summarizes the session cache size (in bytes) required by FreeRADIUS for each algorithm. 
The \gls{PQC} algorithms impose substantially higher storage demands compared to the current RSA based authentication. 
For instance, Falcon-512 requires approximately 85\,\% more storage, while SLH-DSA-256f exhibits an increase of up to 4\,122\,\% relative to RSA-2048. 
Compared to RSA, ML-DSA requires  267\,\%, 395\,\% and 558\,\% more storage for Security Level 1, 3 and 5, respectively.
Among the SLH-DSA variants, the smaller parameter sets require roughly half the storage of their faster counterparts.
Despite these significant increases, the absolute storage requirements remain modest in practical deployments. 
Even in large-scale enterprise environments, the storage overhead remains feasible. 
Ideally, companies should use stateless session resumption. 

Overall, these results indicate that integrating \gls{PQC} into EAP-TLS session resumption introduces only minor performance overheads, confirming its practical feasibility even under varying network conditions.

\textbf{Key Results: }\emph{EAP-TLS session resumption accelerates authentication for all \gls{PQC} algorithms but still remains slightly slower than RSA+ECDH due to larger ML-KEM key share.
Ideally, stateless session resumption should be deployed, as stateful session resumption with \gls{PQC} algorithms requires significantly more storage  with up to over 4\,122\% more, while the absolute requirements remain small, ranging from 50,574 bytes (SLH-DSA-256f) to 2,210 bytes (Falcon512) per session.}

\begin{figure*}
    \centering
    \includegraphics[]{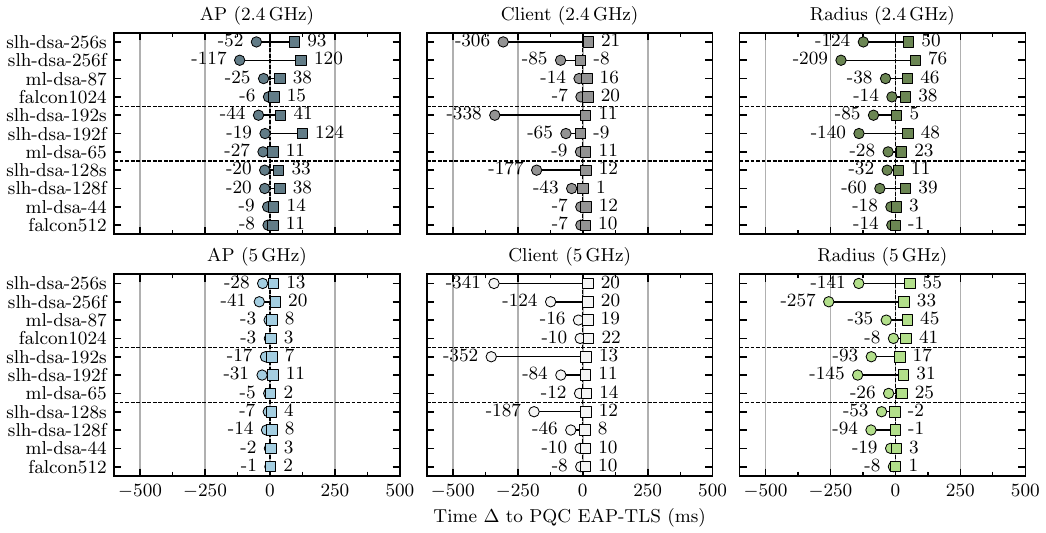}
    \caption{Overview of time delta ($\Delta$) for situation a) between EAP-TTLS left part and the PQC-Hybrid variations right part. The zero baseline indicates the measurements for PQC EAP-TLS. Negative values indicate speed improvements.}
    \label{fig:hybriddelta}
\end{figure*}

\subsection{Hybrid EAP-TLS and EAP-TTLS}
As the deployment of quantum-resistant authentication mechanisms becomes increasingly relevant, hybrid \gls{PQC} offers a practical migration path to secure enterprise networks. 
Hybrid schemes combine a classical cryptographic primitive with a \gls{PQ} counterpart, providing security even if one of the components is later broken. 
This section examines the use of hybrid \gls{PQC} algorithms within EAP-TLS and defines the specific parameter sets used for each security level.
For the evaluation of this work of hybrid PQC within EAP-TLS, we follow the common mapping between NIST security levels and classical key sizes.
For the hybrid \gls{KEM} variation, we use the following configurations: 
\begin{itemize}
    \item Security level 1/2: ECDSA using the p256 curve and KEM X25519MLKEM512.
    \item Security level 3: ECDSA using the p384 curve and KEM X25519MLKEM768.
    \item Security level 5: ECDSA using the p521 curve and KEM SecP384r1MLKEM1024.
\end{itemize}

\gls{EAP-TTLS} authentication requires only the server to present a certificate for authentication.
Consequently, the client only needs to verify the certificate of the server and does not require the transmission of its own certificate, eliminating one sign operation at the client and one verify operation at the server compared to mutual authentication in EAP-TLS.

Figure~\ref{fig:hybriddelta} shows the time delta of \gls{EAP-TTLS} and \gls{PQC}-Hybrid variations compared to the authentication time of pure \gls{PQC} \gls{EAP-TLS} for evaluation situation a).
The left makers represent the \gls{EAP-TTLS} authentication, while the right markers show the \gls{PQC}-Hybrid variations.
The time differences are shown for the three areas (client, AP, and server) at both 2.4\,GHz and 5\,GHz. 

Across all algorithms and security levels, EAP-TTLS achieves faster authentication times at the client, RADIUS server, and AP. 
While this reduction applies to all algorithms, it is larger for both SLH-DSA schemes and for higher security levels. 
The performance increase at the AP is more evident in the 2.4\,GHz band, due to the reduced number of messages resulting from the absence of a client certificate, but still apply to the 5\,GHz band. 

The \gls{PQC}-hybrid algorithms increase the authentication time at both the client and the RADIUS server. 
This increase is more pronounced at higher security levels due to the additional cryptographic operations, including the ECDSA sign/verify steps and the required ECDH key exchange. 
Note that, the resulting overhead affects the absolute authentication time of the faster algorithms (ML-DSA and Falcon) more noticeably than that of the SLH-DSA schemes. 
Since both SLH-DSA variation already have long authentication times, the additional delay introduced by the hybrid \gls{PQC} has only a minor relative impact.
In contrast, Falcon and ML-DSA have faster authentication times, so the hybrid overhead represents a proportionally larger slowdown. 

\textbf{Key Results: } \emph{
PQC-hybrid variants result in longer authentication times, which affects the faster ML-DSA and Falcon algorithms more than the slower SLH-DSA schemes. 
Larger increases are observed at higher security levels.
The EAP-TTLS authentication requires less time at the client, \gls{AP}, and RADIUS server.
This decrease is especially dominant at the client due to the reduced digital signature operations (sign and verify) and data transmission (client certificate).
This decrease is largest for both SLH-DSA variations, while the reduction for all others algorithms remains relative small but still noticeable.}

\section{Discussion}
\label{sec:disc}
In the following section, we discuss the evaluation results, putting them in the context of WPA-Enterprise networks that are considering the deployment of \gls{PQC} algorithms.
While slower than RSA, Falcon and ML-DSA achieve acceptable authentication speeds for enterprise deployments. 
Although security level 1 delivers faster authentication, we recommend using security levels 3 or 5, which provide substantially higher security while maintaining acceptable performance.
Those aiming to deploy variations of SLH-DSA should be mindful of minimal data rates, as higher minimal data rates can significantly reduce authentication time by allowing faster transmission.
Even though the required duration for the authentication is an important metric, it is not the only metric to be considered. 
Equally important are the security level and the number of messages exchanged between the client and server.
Furthermore, minimizing the number of exchanged messages helps to prevent \gls{DoS} attacks against the authentication server and reduces airtime consumption.
\begin{figure}
    \centering
    \includegraphics[]{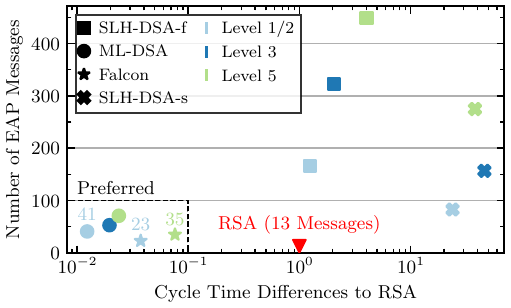}
    \caption{Overview of \gls{PQC} algorithms number of required messages and cycle time differences compared to RSA-2048.}
    \label{fig:alogrithm-comparision}
\end{figure}

Figure~\ref{fig:alogrithm-comparision} summarizes and compares the evaluated \gls{PQC} algorithms based on the number of \gls{EAP} messages and the required cycle time differences compared to classical cryptography, i.e., RSA and elliptic curve cryptography. 
The recommended region highlights algorithms that require fewer than 100 \gls{EAP} messages, which minimizes channel contention and airtime consumption.
Additionally, algorithms in this area require less computational resources compared to RSA, as otherwise they would be susceptible to \gls{DoS} attacks.
Based on this comparison and the previous evaluation, we recommend ML-DSA and Falcon for companies aiming to transition towards \gls{PQC}-secure WPA-Enterprise authentication.
Both algorithms provide fast authentication with slightly more messages than the current combination of RSA and elliptic curve cryptography. 

In contrast, the significantly larger computational requirements of SLH-DSA-s make it vulnerable to \gls{DoS} attacks, in which an attacker maliciously sends excessive amounts of requests overloading the RADIUS server.
Network administrator wanting to utilize SLH-DSA-s should be aware of this implication. 
In comparison, SLH-DSA-f requires significantly more messages which can be problematic if the authentication RADIUS server is located multiple hops away or if client and access point are separated by large distances, resulting in higher latency and airtime consumption.  
Nevertheless, the solid mathematical properties of SLH-DSA make it a secure method and, if used in conjunction with a stateless session resumption, allow for an alternative solution.

Adapting the FreeRADIUS server for \gls{PQ} algorithms is relatively straightforward.
Since OpenSSL already provides implementations of several \gls{PQC} schemes and liboqs can be integrated seamlessly, the overall effort required for integration remains manageable.
This effort encompasses both the integration of liboqs into hostapd and FreeRADIUS and the necessary increase of the default 50-round \gls{EAP} limit, which is insufficient for certain schemes such as SLH-DSA.

It is evident that there is no direct drop-in replacement for RSA-based certificates with respect to latency and memory requirements among the schemes investigated in this work. 
This is expected, as \gls{PQ} schemes typically introduce higher computational costs or larger key sizes compared to established non-\gls{PQC} schemes.

Future work should investigate the impact of larger certificate chains and the additional overhead introduced by certificate revocation lists. 
Furthermore, future work should examine RADSEC (RADIUS over TLS), along with scenarios involving larger and more complex network environments characterized by increased round-trip times.
Additionally, the standardization of \gls{PQC} algorithms is an ongoing process, and other schemes such as MAYO, SQIsign, or MQOM may become standardized in future rounds of the \gls{NIST} additional signature standardization process~\cite{NISTsecondround}. 
These schemes should also be evaluated proactively, even before the additional \gls{NIST} round is completed.

Finally, session resumption proves to be an effective optimization. 
It provides a predictable performance gain and makes it possible to employ even larger algorithms efficiently, without negatively impacting the user experience.
In environments where session resumptions occur frequently and EAP-TLS authentications are comparatively rare, integrating larger algorithms becomes more practical.

\section{Related Work}
\label{sec:related}
The transition to \gls{PQC} is a pivotal element in ensuring the security of wireless network systems against the threats posed by quantum computers in the future.
A a wide range of related research has explored and assessed the integration of \gls{PQC} in various systems.
In the following, we summarize the recent literature and provide an overview of the \gls{PQC} integration methods and challenges.

Bozhok et al.~\cite{bozhkoPerformanceEvaluationQuantumResistant2023} evaluated the performance of TLS with \gls{PQC} algorithms in consumer IoT devices. 
They tried to find the optimal performing \gls{PQC} algorithm in Bluetooth Low Energy (BLE) and Wi-Fi.
Their experimental findings demonstrated that Kyber (ML-KEM), Falcon, and Dilithium (ML-DSA) exhibited the shortest handshake delay.
The authors extend on their work in~\cite{HANNA2025101650} adding the additional overhead of certificate revocation lists and energy consumption of the \gls{PQC} algorithms.
In contrast, we analyze and compare the performance of \gls{PQC} algorithms in different evaluation situations with varying signal quality for WPA-Enterprise networks focusing on RADIUS server, and not specifically on IoT devices.

The work of Bürstinghaus-Steinbach et al.~\cite{burstinghaussteinbachPostQuantumTLSEmbedded2020a}
investigated the feasibility of using \gls{PQC} algorithms on constrained IoT devices by implementing Kyber (ML-KEM) and SPHINCS+ (SLH-DSA) within the mbed TLS library. 
Their study uses the ESP32 microcontroller connected via Wi-Fi, although the performance of the wireless communication layer was not explicitly measured.
The authors primarily focused on the computational overhead introduced by the \gls{PQC} algorithms during the TLS handshake, with SPHINCS+ identified as particularly resource-intensive. 
Memory consumption and code size were also evaluated, with results indicating that despite the increased resource demands, deploying \gls{PQC}-based TLS handshakes on embedded devices is indeed feasible.
Related work have similarly examined the integration of \gls{PQC} mechanisms on resource-constrained and IoT devices, addressing comparable challenges related to limited computational power and memory resources~\cite{10.1007/978-3-031-21280-2_24, 9919545}. 
However, these works predominantly focus on embedded or IoT environments and do not consider the characteristics of enterprise wireless network deployments.
This work focuses explicitly on the integration of \gls{PQC} in enterprise wireless networks.
We examine all the \gls{NIST} standardized \gls{PQC} algorithms and their performance in wireless EAP-TLS and EAP-TTLS communication.

Sosnowski et al.~\cite{sosnowskiPerformancePostQuantumTLS2023} compared the performance of traditional and \gls{PQC} algorithms for TLS. 
Their results show that the performance of ML-KEM is equal compared to current state-of-the-art asymmetric algorithms.
ML-DSA and Falcon are even faster. 
The work of Sikeridis et al.~\cite{sikeridisAssessingOverheadPostquantum2020} examined the overhead of \gls{PQC} in TLS 1.3 and SSH. 
To analyze the performance behavior, they used one local host machine and three remote servers to create a realistic network condition.
They concluded that the integration of \gls{PQC} can lead to an 1-300\% increase in TLS latency, but do not consider EAP-TLS.
Further work has investigated the impact of \gls{PQ} signature schemes in the context of TLS~1.3~\cite{sikeridisPostQuantumAuthenticationTLS2020a, MONTENEGRO2026108062, tzinos2022evaluating, paul2022mixed, 10.1007/978-3-031-49187-0_14, cryptography9040073, 10.1007/978-3-032-08124-7_6}.
Sikeridis et al.~\cite{sikeridisAssessingOverheadPostquantum2020} evaluates authentication using standalone \gls{PQC} algorithms without combining them with classical counterparts.
The authors analyze how these algorithms affect handshake latency, identifying Falcon and Dilithium as the best performing candidates. 
Their results show that the use of \gls{PQ} signatures has a negative impact on the server throughput. 
To mitigate performance overhead, the study explores the use of certificate chains that include multiple \gls{PQ} algorithms, which helps improve overall handshake time and throughput.
In this work, we examine the performance of \gls{PQC} in TLS within wireless networks. 

Cho et al.~\cite{PQC-MACsec} propose an authenticated \gls{PQ} key establishment protocol to enable quantum-resistant MACsec and MACsec Key Agreement protocol.
This is achieved by authenticating and exchanging the master session key using a radius server and \gls{PQ} EAP.
In contrast to this work, the authors concentrate in their work on the quantum resistance of MACsec.
We are investigating the impact of \gls{PQC} on enterprise wireless networks in terms of performance and latency.

None of the existing research analyses the impact of \gls{PQC} on WPA-Enterprise authentication.
To the best of our knowledge, we are the first to quantify the duration of EAP-TLS and EAP-TTLS in real-world situations, showing the feasibility of \gls{PQC} in WPA-Enterprise authentication, effectively protecting against \gls{PQ} attacks.

\section{Conclusion}
\label{sec:conclusion}
\glsresetall
Quantum computers pose a significant threat to many application domains, including wireless communication systems such as WPA-Enterprise networks. 
To mitigate this risk, it is essential to investigate how these systems can adopt \gls{PQ} secure algorithms.
This work assesses and quantifies the real-world impact of \gls{PQC} on WPA-Enterprise authentication.
We show that a \gls{PQ}-secure WPA-Enterprise authentication is possible with common open-source projects, namely hostap and FreeRADIUS.
Additionally, we analyze the duration of a \gls{PQ}-secure WPA-Enterprise authentication inside an office environment with three evaluation situations of varying signal quality: a) excellent signal quality, b) good signal quality, and c) poor signal quality.
We show that Falcon and ML-DSA perform similarly across all evaluation situations compared to RSA and therefore are suitable replacements in \gls{PQ}-secure solutions. 
Based on these findings, we recommend the use of Falcon at security level 5 or ML-DSA at security 3. 
The SLH-DSA-s variants are better suited for scenarios with weak signal quality than the SLH-DSA-f variants, as they generate fewer packets and thereby reduce transmission overhead. 
However, both variants still result in significantly longer authentication times compared to ML-DSA and Falcon. 
Nevertheless, the additional authentication overhead is unlikely to pose a problem for most organizations, especially when session resumption is employed. 

In summary, this work shows that WPA-Enterprise networks can be secured against quantum-capable adversaries by integrating suitable \gls{PQ} algorithms, paving the way for a smooth and practical migration towards \gls{PQ}-secure wireless communication.

\bibliographystyle{IEEEtran}
\bibliography{paper}

\begin{IEEEbiography}[{\includegraphics[width=1in,height=1.25in,clip,keepaspectratio]{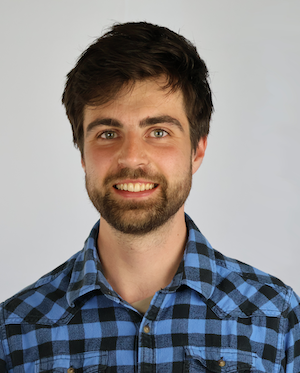}}]{LUKAS KÖDER } received the M.S. degrees in Applied Informatics with a specialization in IT Security in 2024. He is currently pursuing the Ph.D. degree in Informatics at Esslingen University of Applied Sciences under the supervision of Prof. Dr. Tobias Heer.

He works as a research associate in the Smart Factory Grids project at the Esslingen University of Applied Sciences, which aims to improve the security and robustness of next-generation industrial networks. His research interests include wireless communication, particularly Wi-Fi, industrial network security, and post-quantum cryptography.
\end{IEEEbiography}

\begin{IEEEbiography}[{\includegraphics[width=1in,height=1.25in,clip,keepaspectratio]{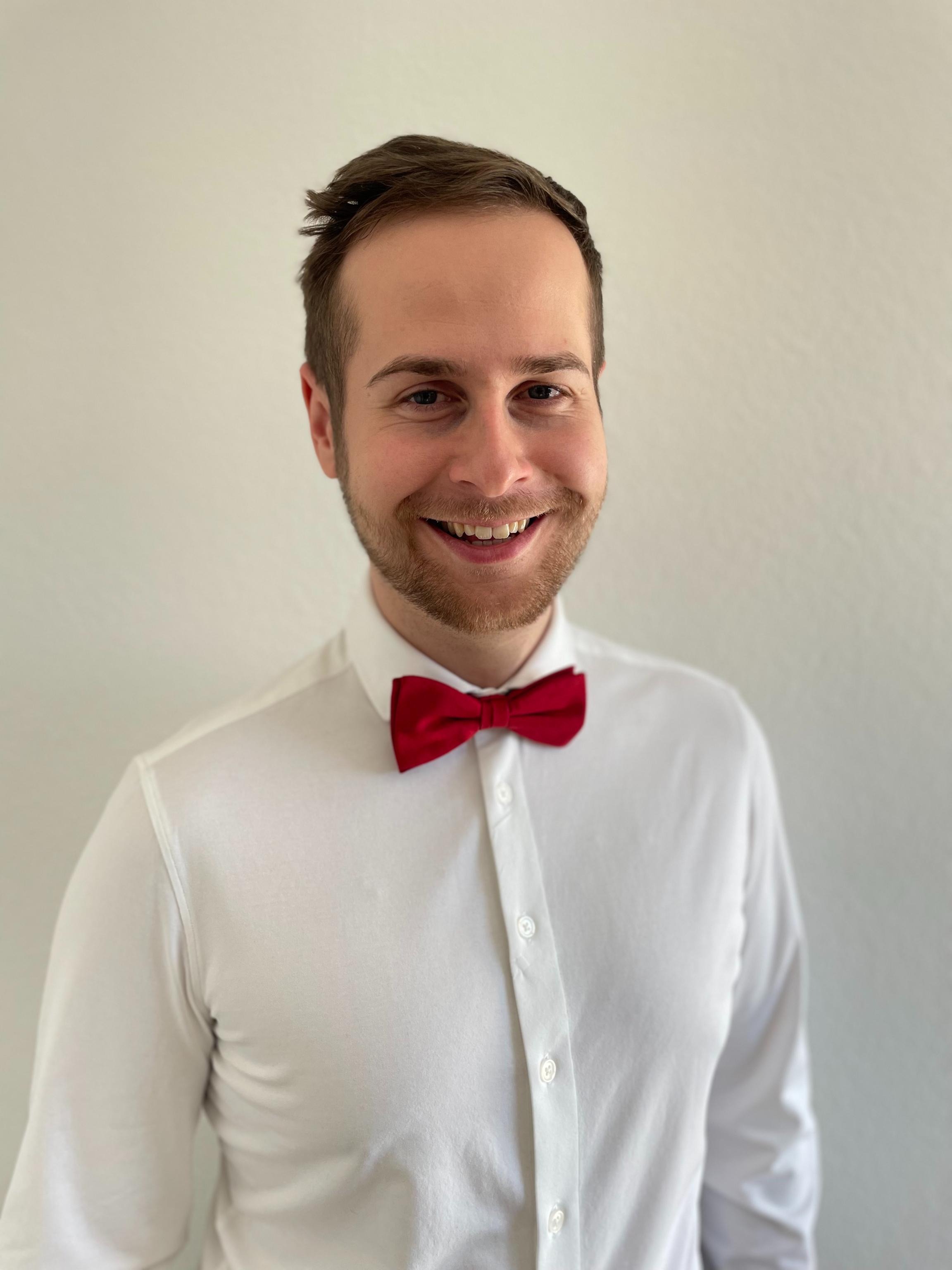}}]{NILS LOHMILLER } 
received the master’s degree, in 2022. 
He is currently pursuing the Ph.D. degree at Esslingen University with Prof. Dr. Tobias Heer and the Chair of Communication Networks of Prof. Dr. Habil. Michael Menth, University of Tübingen, Germany.
His research interests include post-quantum cryptography, automotive security, and Internet of Things (IoT).
\end{IEEEbiography}

\begin{IEEEbiography}[{\includegraphics[width=1in,height=1.25in,clip,keepaspectratio]{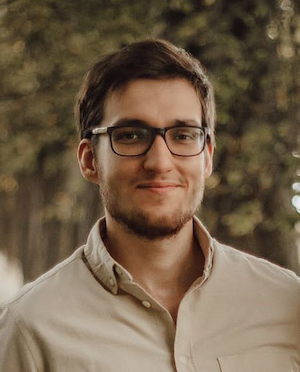}}]
{PHIL SCHMIEDER } received his master’s degree in computer science in late 2024. He is currently pursuing a Ph.D. degree at Wuppertal University with Prof. Dr. Tibor Jager at the Chair of IT Security and Cryptography. His research interests are the post-quantum transition of Public Key Infrastructure, hybrid cryptography and efficient special-purpose cryptographic schemes.
\end{IEEEbiography}

\begin{IEEEbiography}[{\includegraphics[width=1in,height=1.25in,clip,keepaspectratio]{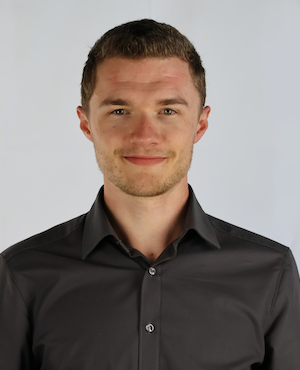}}]{BASTIAN BUCK } received the master’s degree from Albstadt-Sigmaringen University, in 2023. He is currently pursuing the Ph.D. degree at Esslingen University with Prof. Dr. Tobias Heer. His research interests include e-mail security, post-quantum cryptography, and network security.
\end{IEEEbiography}

\begin{IEEEbiography}[{\includegraphics[width=1in,height=1.25in,clip,keepaspectratio]{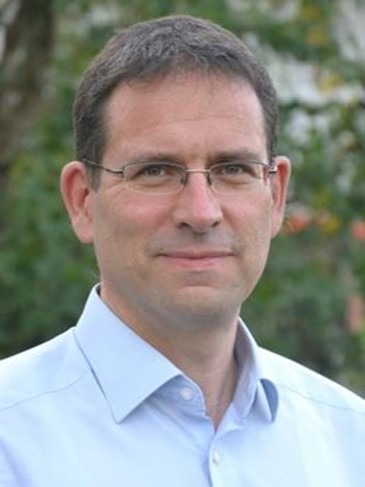}}]{MICHAEL MENTH } (Senior Member, IEEE)  received the Diploma degree from The University of Texas at Austin, Austin, TX, USA, in 1998, the Ph.D. degree from Ulm University, Germany, in 2004, and the Habilitation degree from the University of Würzburg, Germany, in 2010. He has been the Chair Holder of Communication Net- works, since 2010. He is currently a Professor with the Department of Computer Science, University of Tübingen, Germany. His special interests
are performance analysis and optimization of communication networks, resilience and routing issues, as well as resource and congestion management. His recent research interests include network softwarization, in particular P4-based data plane programming, time-sensitive networking (TSN), the Internet of Things, and internet protocols. He contributes to standardization bodies, mainly to the IETF.
\end{IEEEbiography}
\begin{IEEEbiography}[{\includegraphics[width=1in,height=1.25in,clip,keepaspectratio]{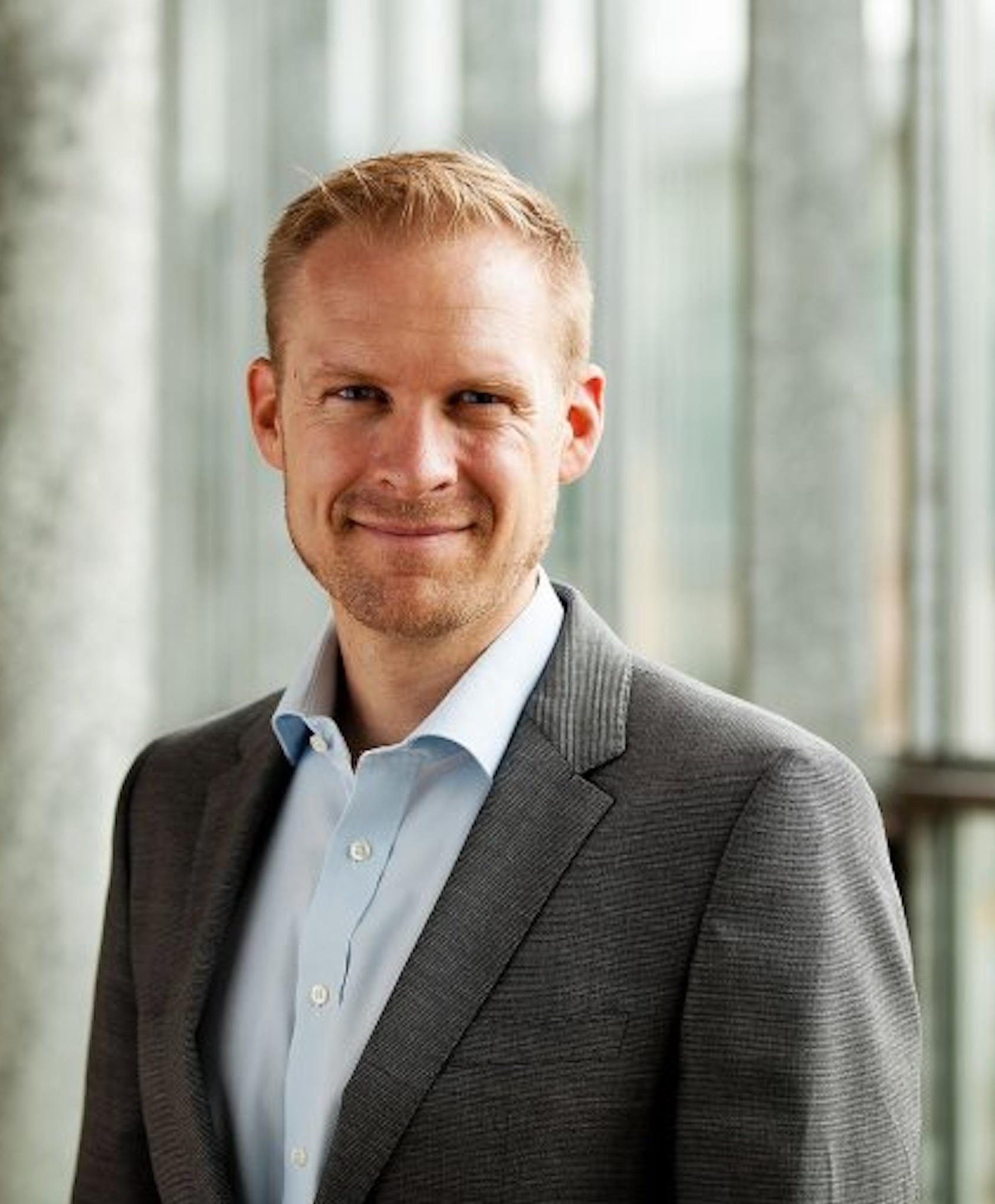}}]{TOBIAS HEER }
		is the Dean of the faculty of Computer Science and Engineering and a Professor at Esslingen University.
		He received his Diploma degree from University of Tuebingen, Germany, in 2006, and the Ph.D degree from RWTH Aachen University, Germany, in 2011.
		His research focus is IT security, network security, and industrial security.
\end{IEEEbiography}

\end{document}